\documentclass[
aps,%
prl,%
twocolumn,%
groupedaddress,%
superscriptaddress,%
showpacs,%
letterpaper,%
amsfonts,%
footinbib,%
10pt,
floatfix,%
noeprint%
]{revtex4-1} 
\RequirePackage{amsmath,amssymb,bm}
\usepackage{wasysym}
\usepackage{float}
\usepackage{graphicx}
\usepackage[dvipsnames]{xcolor}
\usepackage[caption=false]{subfig}
\usepackage{lineno}
\usepackage{ulem}

\begin{document}

\title{Antiferromagnetism and chiral d-wave superconductivity from 
an effective $t-J-D$ model for 
twisted bilayer graphene}
\author{Xingyu Gu}
\affiliation{Department of Physics, National University of Singapore, 117542, Singapore}
\affiliation{Centre for Advanced 2D Materials and Graphene Research Centre, National University of Singapore, 117546, Singapore}

\author{Chuan Chen}
\affiliation{Department of Physics, National University of Singapore, 117542, Singapore}
\affiliation{Centre for Advanced 2D Materials and Graphene Research Centre, National University of Singapore, 117546, Singapore}

\author{Jia Ning Leaw}
\affiliation{Department of Physics, National University of Singapore, 117542, Singapore}
\affiliation{Centre for Advanced 2D Materials and Graphene Research Centre, National University of Singapore, 117546, Singapore}

\author{Evan Laksono}
\affiliation{Department of Physics, National University of Singapore, 117542, Singapore}
\affiliation{Centre for Advanced 2D Materials and Graphene Research Centre, National University of Singapore, 117546, Singapore}

\author{Vitor M. Pereira}
\affiliation{Department of Physics, National University of Singapore, 117542, Singapore}
\affiliation{Centre for Advanced 2D Materials and Graphene Research Centre, National University of Singapore, 117546, Singapore}

\author{Giovanni Vignale}
\affiliation{Centre for Advanced 2D Materials and Graphene Research Centre, National University of Singapore, 117546, Singapore}
\affiliation{Yale-NUS College, 16 College Avenue West, 138527, Singapore}
\affiliation{Department of Physics and Astronomy, University of Missouri, 65201, USA}

\author{Shaffique Adam}
\affiliation{Department of Physics, National University of Singapore, 117542, Singapore}
\affiliation{Centre for Advanced 2D Materials and Graphene Research Centre, National University of Singapore, 117546, Singapore}
\affiliation{Yale-NUS College, 16 College Avenue West, 138527, Singapore}

\begin{abstract}
Starting from the strong-coupling limit of an extended Hubbard model, we develop a spin-fermion theory to study the insulating phase and pairing symmetry of the superconducting phase in twisted bilayer graphene.
Assuming that the insulating phase is an anti-ferromagnetic insulator, we show that 
fluctuations of the anti-ferromagnetic order in the conducting phase can mediate superconducting pairing. Using a self-consistent mean-field analysis, we find that the pairing wave function 
has a chiral d-wave symmetry.  Consistent with this observation, we show explicitly the existence of chiral Majorana edge modes by diagonalizing our proposed Hamiltonian on a finite-sized system.
These results establish twisted bilayer graphene as a promising
platform to realize topological superconductivity. 
\end{abstract}
\maketitle

\emph{Introduction}.
Correlated insulating phases and unconventional
superconductivity have been recently observed in twisted bilayer
graphene (tBG) near the magic twist angle $\theta\approx 1.1^{\circ}$~\cite{cao2018correlated,cao2018unconventional,yankowitz2019tuning}.  This is currently leading to an intense surge of interest partly due to the similarity of its phase diagram and that of high-$T_c$ 
superconductors. Since the experimental discovery,
the nature of the insulating phase and pairing mechanism in tBG have been
studied by several theoretical proposals,
that start from either a weak or strong coupling limit~\cite{xu2018topological,isobe2018superconductivity,liu2018chiral,laksono2018singlet, yuan2018model,you2018superconductivity,po2018origin,koshino2018maximally,kang2018symmetry,guo2018pairing,huang2018antiferromagnetically,dodaro2018phases,xu2018Kekule,zou2018band,fidrysiak2018unconventional,wu2018coupled}.  From the weak coupling point of view, the interaction is included
perturbatively to the free-electron band structure, calculated from a continuum model~\cite{bistritzer2011moire,dos2007graphene}.
Because of the the van-Hove singularity and Fermi surface nesting~\cite{isobe2018superconductivity,liu2018chiral, you2018superconductivity,laksono2018singlet}, the system exhibits strong particle-hole fluctuations that can lead to an insulating
phase and superconductivity upon doping. 

\begin{figure}[h!]
\begin{centering}
\includegraphics[width=0.45\textwidth]{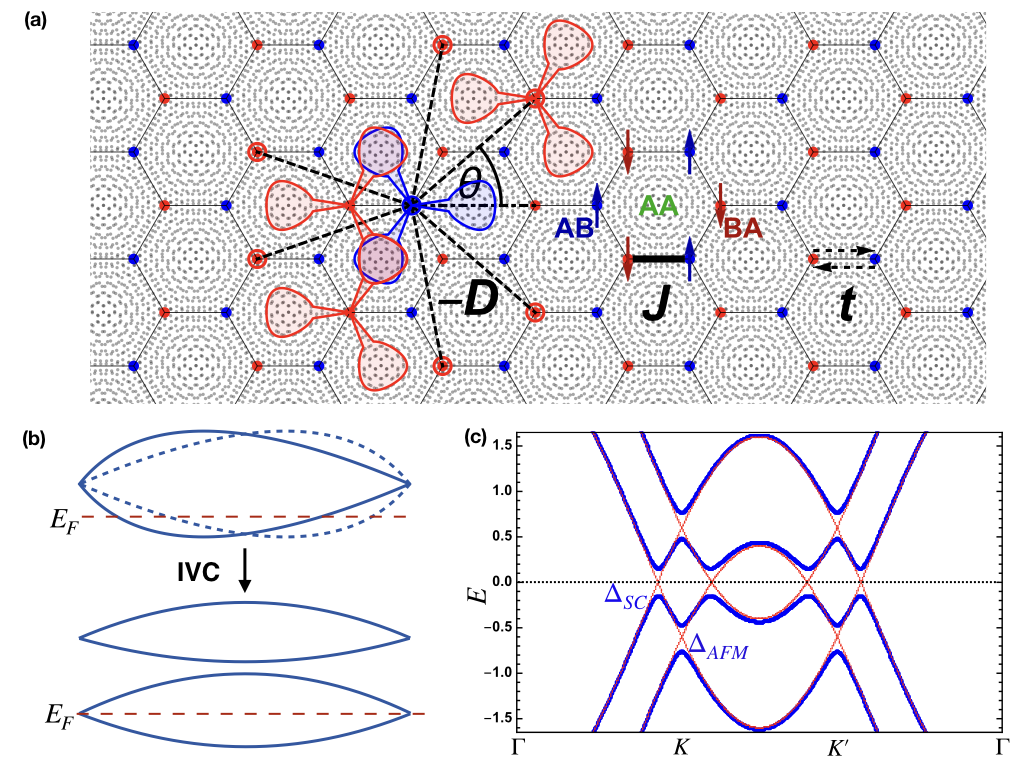}
\caption{(a) The underlying moir\'e pattern and the emergent honeycomb lattice. The lobes show the unusual shape of Wannier functions schematically. The three red Wannier functions share 0, 2 and 1 peaks with the blue Wannier function from top to bottom.  The hopping, antiferromagnetic coupling and pairing are denoted as $t$, $J$ and $-D$, respectively, and the arrows represent the short-range antiferromagnetic order. (b) The occurrence of intervalley coherence order shifts the Fermi level from half-filling to the Dirac point. (c) Gap opening at finite chemical potential. The red line is the non-interacting spectrum and the blue line is the spectrum with non-zero AFM and superconducting gap. The SC gap is at Fermi level (indicated by the dashed line) while AFM gap is at Dirac point. The $k$ path is along the $\Gamma-K-K'-\Gamma$ line in the moir\'e brillouin zone.}
\label{fig:fig1}
\end{centering}
\end{figure}

On the other hand, the multiple insulating phases observed in the experiment~\cite{yankowitz2019tuning} could justify the strong coupling limit as more appropriate for the reference state.  Xu and Balents proposed a $SU(4)$ Hubbard model on a triangular lattice motivated by the fact that the density of electrons in tBG peaks at the AA regions~\cite{xu2018topological}. However, it was later shown that to correctly incorporate the moir\'e symmetries, one needs to include at least two orbitals in the underlying tight-binding Hamiltonian~\cite{isobe2018superconductivity,liu2018chiral, you2018superconductivity,laksono2018singlet,kang2018symmetry,yuan2018model,koshino2018maximally,po2018origin,po2018origin}.  More recently, by taking into account the unusual shape of the Wannier functions, an extended Hubbard model with a ``cluster charge interaction'' between the electrons (discussed below) was proposed~\cite{po2018origin,koshino2018maximally}, where the Mott insulating phase and reduced Landau level degeneracy at half-filling are explained by an ``inter valley coherence'' (IVC) (where the monolayer valley degree of freedom is polarized in xy plane), and the chemical potential at the half-filled moir\'e band is shifted to the Dirac point. The shifting of Fermi level are schematically shown in Fig. 1(b). This loss of monolayer valley quantum number explains why the experimentally observed Landau level degeneracy close to half filling is half that at charge neutrality.  This breaking of valley symmetry (as opposed to spin symmetry) is supported by Hartree-Fock calculations~\cite{po2018origin}, and because experiments suggests spin-singlet Mott-like behaviour~\cite{cao2018correlated}, which would require unpolarized spins.  The IVC state would still be semi-metallic and, to obtain an insulating phase, further symmetry breaking must take place to gap the Dirac cone.

Reference~\cite{xu2018Kekule} explored the possible symmetry-broken phases using quantum Monte Carlo calculations for such a cluster charge interaction and, as in monolayer graphene when the contact interaction dominates over the long-range Coulomb interaction, the ground state in the strong coupling limit has anti-ferromagnetic (AFM) order~\cite{tang2018role}.  Although the AFM insulating phase in this model has the same origin as in the monolayer, the cluster charge interaction reduces the effective on-site repulsion, and this might explain the small gap of the insulating phase observed experimentally. Unlike monolayer graphene, the Monte Carlo results also found a Kekul\'e valence bond solid phase at intermediate coupling, but this is not germane to our considerations here.

In this Letter, we start from the above-mentioned strong coupling AFM limit with assumed intervalley coherence order.  Due to the valley $U(1)$ symmetry breaking, there is no topological obstruction to build up a tight binding model~\cite{zou2018band}. We show that the low energy physics is governed by a $t-J-D$ model, $H=H_{t}+H_{J}+H_{D}$, where 
\begin{subequations}
\begin{align}
&H_t=-t \sum_{\langle ij \rangle, \alpha}
a^{\dagger}_{i\alpha}b_{j\alpha}+h.c.\\
&H_J=J\sum_{\langle ij \rangle} \Vec{S}_{i}\cdot \Vec{S}_{j}\\
&H_D=-D\sum_{\langle \langle ij \rangle \rangle} h^{\dagger}_{ij}h_{ij} \label{eq:Hv}
\end{align}
\label{eq:H1}
\end{subequations}

\noindent Here, $t$ is hopping parameter, $i$, $j$ are sites of the honeycomb lattice defined by the AB, BA regions in the moir\'e lattice, $a$ and $b$ are the electron operators corresponding to two sublattices, $\alpha$ indicates
spin, 
$\langle \langle ij \rangle \rangle$ in Eq. (\ref{eq:Hv}) represents
the blue and red pairs in Fig. \ref{fig:fig1}a. $D$ is the pairing strength, 
$h_{ij}=a_{i\uparrow}b_{j\downarrow}-a_{i\downarrow}b_{j\uparrow}$
is the inter-sublattice singlet operator. $J$ is AFM coupling and $\Vec{S}$ is the spin operator. 
Note that, unlike in the approach to cuprates or pnictides~\cite{hu2016identifying}, in this model the superconducting pairing couples not the nearest but remote neighbors, as schematically shown in Fig. 1(a).

Based on this Hamiltonian, our main conclusion is that the pairing symmetry in tBG should be chiral $d+id$ (this is consistent with findings using other models used for tBG, including Refs.\cite{xu2018topological,guo2018pairing,huang2018antiferromagnetically,liu2018chiral,fidrysiak2018unconventional,wu2018coupled}).  The advantage of our explicit Hamiltonian is that we can directly demonstrate that quantum fluctuations of the AFM order mediate an attractive pairing interaction.
Moreover, since we have the form of that pairing in real space, using a finite size geometry, we can prove the existence of chiral Majorana modes that should be observable in experiments.

\emph{Effective model}. To derive Eq. (\ref{eq:H1}), we start from the following real-space Hamiltonian $H=H_{t}+H_{U}$ on a honeycomb lattice \cite{po2018origin,koshino2018maximally}, where
\begin{subequations}
\begin{align}
H_t &= -t \sum_{<ij>,\alpha}a^{\dagger}_{i\alpha}b_{j\alpha}
+ h.c.\\
H_U &= U\sum_{R}(Q_R-2)^{2}
\end{align}
\end{subequations}
Here $R$ marks the position of each hexagon's center, and 
$Q_{R}=\sum_{i \in \hexagon} \sum_{\alpha}\frac{n_{i\alpha}}{3}$
is the charge located at that position.

Compared to the onsite Hubbard interaction, there is repulsive Coulomb interaction between electrons sharing the the hexagon in this model. This unusual form of interaction originates from the extended nature of the Wannier functions in tBG. For the flat bands in tBG, it is known that the associated electron density peaks form a triangular lattice~\cite{cao2018correlated}. In contrast, symmetry dictates that the Wannier centers span a honeycomb lattice~\cite{kang2018symmetry,yuan2018model,koshino2018maximally,po2018origin}. The system accommodates these constraints by exhibiting Wannier functions with a three-lobed shape~\cite{kang2018symmetry,koshino2018maximally}. While these are centered at the honeycomb sites, the spatial superposition of the Wannier functions at the corners of an hexagon creates maxima of the electronic density at its center. As a result, there is overlap of 3, 2, 1 peak(s) when considering the interaction between on-site, NN, next NN (NNN) and third NN electrons. This determines the interaction strength ratio from onsite to third NN to be 3:2:1:1. The Wannier functions are schematically shown in Fig. 1(a).

Since the interaction is larger on-site than among NN, an anti-ferromagnetic coupling $J \sim t^2/U$
arises to leading order in perturbation theory~\cite{anderson1950antiferromagnetism} (explaining the numerical discovery of the AFM order in the quantum Monte Carlo~\cite{xu2018Kekule}).  We note in passing that this AFM insulating phase is also consistent with
the experimentally observed quantum oscillation
data~\cite{cao2018correlated,cao2018unconventional,yankowitz2019tuning} as follows: 
In the IVC theory, considering the spin and moir\'e valley degree of freedom, the Fermi surface should be four-fold degenerate around half filling, but it is observed only 2-fold degenerate in  experiment. This reduction can be understood as the
splitting of Landau levels, as has been discussed previously~\cite{choi2011angle,de2011topologically,lee2011quantum,moon2012energy}. This argument also explains the anomalous FS degeneracy reduction around charge neutrality, compared to bilayer graphene with a larger twist angle $\theta=1.8^{\circ}$~\cite{cao2016superlattice}.

In cuprates and iron pnictides, the AFM coupling is closely related to
superconductivity~\cite{hu2016identifying}, which can be understood
using the Fierz identity:
$\Vec{S}_{i}\cdot \Vec{S}_{j}=-2h^{\dagger}_{ij}h_{ij}+n_{i}n_{j}$,
where $h_{ij}$ is the singlet pairing operator defined above.
But in tBG, one expects the NN pairing to be strongly suppressed due to the strong repulsion between electrons in the same hexagon, similarly to how the on-site pairing is suppressed in cuprates and pnictides.
Therefore, to determine the nature and feasibility of a superconducting ground state, to leading order we need only consider the pairing between electrons which not sharing a common hexagon, because only these are weakly repelled from each other.
In our study, we consider the six nearest ones indicated by the
red points in Fig. \ref{fig:fig1}a and we show that such long range
pairing can be mediated by the AFM fluctuations.
First we sketch the basic physics of AFM mediated SC, a more concrete derivation is shown in supplymentary S1. 
The coupling between AFM order parameter $m_{i}=\frac{\langle S^{z}_{a}\rangle - \langle S^{z}_{b}\rangle}{2}$ and the fermions is:
\begin{equation}
H_{i}=\lambda m_{i}\sigma^{z}_{\alpha \beta} (a^{\dagger}_{i\alpha}a_{i\beta}
-b^{\dagger}_{i\alpha}b_{i\beta}),
\label{eq:coupling1}
\end{equation}
where $\lambda$ is the coupling strength. 
$a$ and $b$ are the electron annihilation operators of the two
sublattices. The local magnetization is assumed to be aligned in the
$z$-direction. 
Here we assume there is no long range AFM order, only the fluctuation is important. The inter-sublattice effective electron-electron
interaction can be obtained by integrating out the spin fluctuations, and reads
\begin{equation}
H_{ij}=\lambda^{2}
\chi_{i,j}
\sigma^{z}_{\alpha \beta} \sigma^{z}_{\gamma \delta} a^{\dagger}_{i\alpha}a_{i\beta} b^{\dagger}_{j\gamma}b_{j\delta}
\label{eq:coupling2},
\end{equation}
It arises in an approximation that neglects retardation effects, where $\chi_{ij}$ is the static spin susceptibility $\chi_{ij}\propto \frac{1}{g} e^{-R_{ij} / \xi}$, with $R_{ij}$ the distance between unit cells $i$ and $j$, $\xi$ the magnetic correlation length, and $g$ the spin stiffness (see supplementary information for details). 
It is easy to see that $H_{ij}$ is attractive for anti-parallel
spins, in which case $\alpha = \beta = -\gamma = -\delta$. The attractive interaction
$D \sim \frac{\lambda^2}{g} e^{-R_{ij} / \xi}$ is significant as long as  spin fluctuation are strong ($\xi$ is large) even though the AFM
order itself doesn't develop.
This attractive interaction between anti-parallel spins leads to singlet
superconductivity~\cite{scalapino1987fermi}, which will be our focus here
since it was furthermore suggested by the experimental results~\cite{cao2018unconventional}. 

Moreover, using similar arguments, one finds that the intra-sublattice interaction
between anti-parallel spin is repulsive.
Such a staggered ``attractive-repulsive-attractive'' profile is similar
to cuprates, where it arises from Fermi surface nesting
\cite{scalapino1987fermi,vignale1989superconductiviting,
vignale1990motion}.
This similarity between pairing mechanism provides some unification between tBG and cuprates. In this work, for simplicity, we only consider the nearest pairing possibilities.
In the following, we study the ground state of the Hamiltonian in Eq.(\ref{eq:H1}) consisting of the kinetic
energy, anti-ferromagnetic coupling between NN sites and a singlet pairing term.

\begin{table}
\begin{center}
\caption{Order parameter form}
\begin{tabular}{| c | c | c | c | c |} 
\hline
Pairing symmetry & $s$ & $d_{x^2-y^2}$ & $d_{xy}$ &
${d_{x^2-y^2}+id_{xy}}$ \\ \hline
form of $f(\theta)$ & 1 & $\cos(2\theta)$ & $\sin(2\theta)$ &
$e^{i2\theta}$ \\ \hline 
\end{tabular} 
\label{tab:op}
\end{center}
\end{table}

\begin{figure}[htbp]
\begin{centering}
\includegraphics[width=0.45\textwidth]{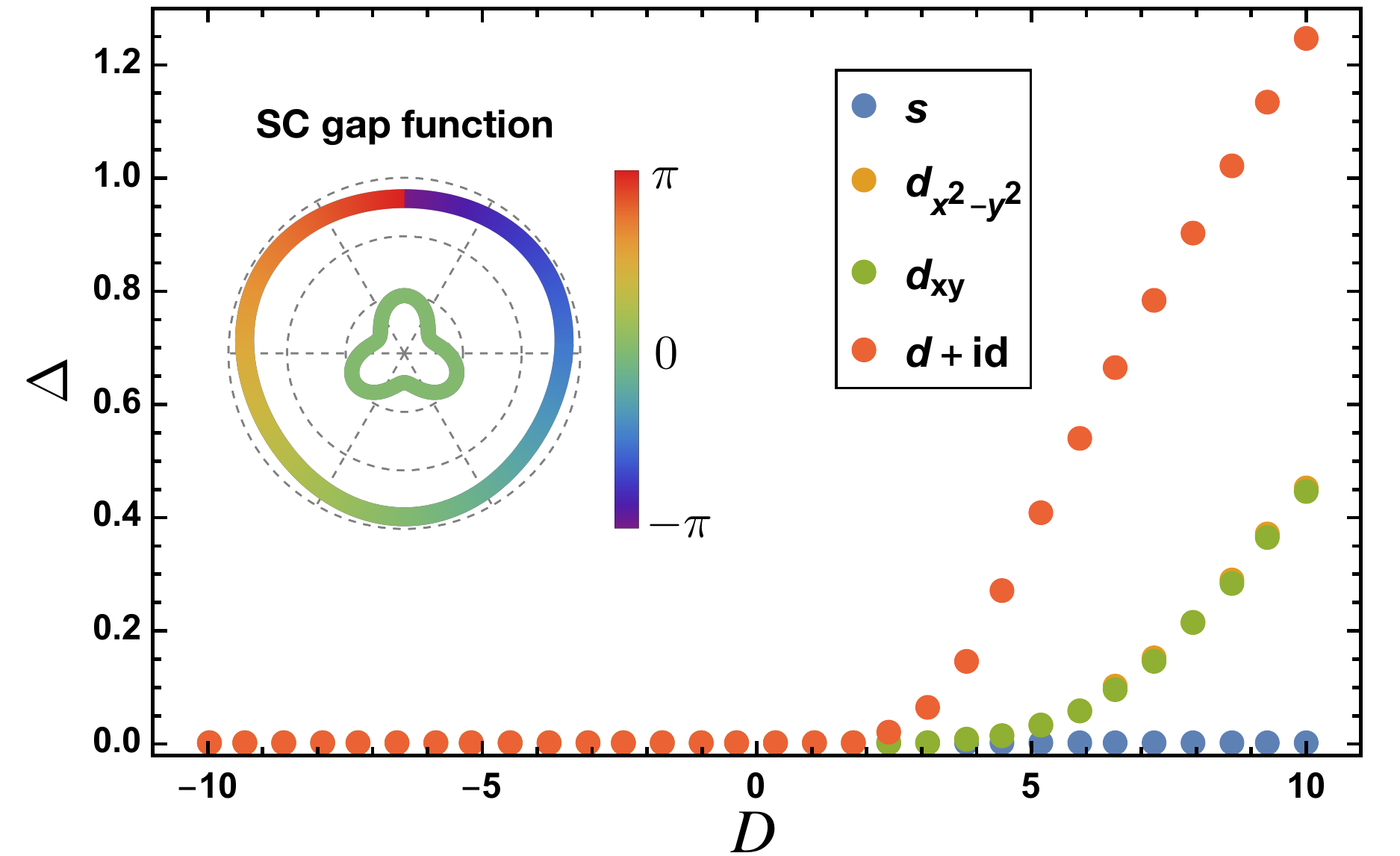}
\caption{Solution of gap equation (\ref{eq:GapEq_sc}) for different
pairing symmetries. The $s$-wave order parameter is always very small,
while the latter three pairing symmetries develop finite order
parameter at finite $D$. The solution for $d_{x^2-y^2}$ in and
$d_{xy}$ are exactly degenerate, which is protected by symmetry.
The $D_c$ of $d+id$ is the smallest, indicating it to be the ground
state. The inset shows the magnitude and phase of the $s$- and
$d+id$-wave gap functions along a circle centered at the Dirac point with
radius $0.1/a$ when real space order parameter $\Delta$ is fixed to be 1.
The radius of the ring shows magnitude and the color shows phase.
The outer ring represents $d+id$ and the inner ring represents $s$-wave.
For clarity, the magnitude of the $s$-wave order parameter has been
increased by a factor of five.
}
\label{fig:fig2}
\end{centering}
\end{figure}
\emph{Pairing symmetry}.
Before studying the interplay between AFM and SC phases, we tackle the SC instability first (with $J=0$) and fix the pairing symmetry. 
The order parameter in real space is defined as $\Delta_{ij}=D\langle h_{ij} \rangle$. The mean field interaction reads
\begin{align}
H_{D}^{MF} &= \sum_{i,\mathbf{a}} -\Delta_{i,\mathbf{a}}h^{\dagger}_{i,i+\mathbf{a}} - \Delta_{i,\mathbf{a}}^{*}h_{i,i+\mathbf{a}} + \frac{6N}{D} |\Delta|^2 \nonumber  \\
&\cong -\sum_{\mathbf{k}} D_{\mathbf{k}}c_{\mathbf{k}\uparrow}^{\dagger}c_{-\mathbf{k}\downarrow}^{\dagger}+h.c.+\frac{6N}{D}|\Delta|^2,
\label{Hscmf}
\end{align}
where the $\Delta_{i,\mathbf{a}}=\Delta_{i,i+\mathbf{a}}$ denotes the
pairing between site $i$ and $i+\mathbf{a}$, with a representing the
six vectors connecting the blue and the red sites.
In the second line, $c$ is the conduction fermion annilation operator,
$D_{\mathbf{k}}=\sum_{\mathbf{a}}
\Delta_{\mathbf{a}}\cos(\mathbf{k}\cdot \mathbf{a}-\phi_{\mathbf{k}})$
is the SC gap. In the expression of $D_{\mathbf{k}}$, $\Delta_{\mathbf{a}}$ are the SC order parameters
defined on the six bonds and $\phi _{\mathbf{k}} =
\arg(\sum_{\boldsymbol{\delta}}e^{i\mathbf{k}\cdot\boldsymbol{\delta}}$),
where $\boldsymbol{\delta}$ are the three vectors connecting the NN
sites. To get this result, we have replaced the electron operators in the sublattice representation
by electron operators in the band representation and keep only the intra-conduction band pairing term,
since it is the most important term at finite chemical potential where SC is observed experimentally (The detailed derivation of the mean
field Hamiltonian Eq.(\ref{Hscmf}) is shown in supplementary section
S-II). 

We consider the four different pairing symmetries indicated in Table. \ref{tab:op}: $s$-wave,
$d_{x^2-y^2}$-wave, $d_{xy}$-wave and ${d_{x^2-y^2}+id_{xy}}$-wave.
They are defined by $\Delta_{\mathbf{a}}=\Delta f(\theta_{\mathbf{a}})$,
with $\theta_{\mathbf{a}}$ the angle
between the bonds and the 
+x-axis, as shown in Fig. \ref{fig:fig1}a.

The pairing symmetry in the ground state has been obtained by a self-consistent mean field analysis.
The spectrum of the Bogoliubov-de Gennes (BdG) Hamiltonian is (See supplementary section S-IV for more details):
$E_{\mathbf{k}}=\pm\sqrt{(\epsilon_{\mathbf{k}}-\mu)^2+|D_\mathbf{k}|^2}$
and $\epsilon_{\mathbf{k}}=t\sqrt{3+2\cos(\sqrt{3}k_y) + 
4\cos(\frac{\sqrt{3}k_y}{2})\cos(\frac{3k_x}{2})}$ is the energy dispersion of free electrons in the honeycomb lattice \cite{neto2009electronic}.
By minimizing the free energy, we obtain the gap equation:
\begin{equation}
\sum_{\mathbf{k}}\frac{1}{N}\frac{|\sum_{\mathbf{a}}f(\theta_{\mathbf{a}})|^2}{E_{\mathbf{k}}}=\frac{12}{D},
\label{eq:GapEq_sc}
\end{equation}
where $N$ is the number of unit cells. 
The solutions of this gap equation as a function of $D$ for different
pairing symmetries are shown in Fig. \ref{fig:fig2}. We choose $\mu=0.2$
for illustration and use the hopping t as the unit of energy.
For repulsive interaction ($D<0$), there is obviously no
superconductivity at all.
For attractive interactions ($D>0$), while the s-wave order parameter remains
zero for $D$ up to 10,  a nonzero d-wave order parameter emerges
at a finite critical interaction $D_c$.
The two different d-wave phases are exactly degenerate, which is
protected by symmetry. The case $d+id$ has both the smallest $D_{c}$ and larger order parameter magnitude at given $D$.
This implies the ground state should be $d+id$.
In tBG, due to the small bandwidth, $t$ is extremely small~\cite{koshino2018maximally} and the attractive interaction strength
can exceed $D_c$ easily. We note that a similar $d+id$ SC phase is obtained in monolayer graphene~\cite{black2014chiral}, if one considers only NN pairing. Hence, $d+id$ pairing s likely a robust feature of correlation-driven SC in honeycomb lattices in the strong coupling limit. But,crucially, whereas in monolayer graphene the interaction strength is too weak ($\frac{U}{t}\sim3.3$~\cite{wehling2011strength,tang2015interaction}) while, in tBG, the flat band drives the system into strong coupling ($\frac{U}{t}\sim50$~\cite{koshino2018maximally}), making it an ideal platform to realize topological superconductivity.

In order to understand the favouring of $d+id$ pairing symmetry,
we study the different gap functions in momentum space with
the real-space amplitudes $\Delta_{i,\mathbf{a}}$ fixed to be unit and compare the behavior of $D_{\mathbf{k}}$.
Since the reduction of free energy is dominated by the gap opening at the Fermi surface,
we plot the SC gap along the circle centered at the Dirac point
with radius $0.1/a$, where $a$ is the lattice constant of the emergent honeycomb lattice.
The cases of $s$-wave and $d+id$ are shown in the inset of Fig.
\ref{fig:fig2}.
The gap due to the former is significantly smaller than that arising from
d-wave pairing, which means d-wave is more favourable in energy. We notice that the same argument have also been used to determine the pairing symmetry in cuprates and iron pnictides~\cite{hu2016identifying}. This provides another opportunity to unify
tBG and high-$T_c$ superconductors.
In addition, we find that while the s-wave order parameter has a trivial phase (given by the color scale in the inset of Fig. 2), the d+id case displays a nontrivial winding phase of 2$\pi$, making it qualitatively similar to the case of $p+ip$ symmetry in monolayer graphene.~\cite{black2015topological}.
%

\emph{Interplay between AFM and SC}.
We now reinstate the AFM interaction ($J\ne0$) and probe the relative stability of the two phases according to the full Hamiltonian in Eq. (\ref{eq:H1}).
According to the discussion above, we consider here a $d+id$ 
superconducting pairing symmetry.
The spectrum of the BdG Hamiltonian is:
$E_{\mathbf{k}}=\pm\sqrt{\epsilon_{\mathbf{k}}^2 + m^2 + \mu^{2} +
|D_\mathbf{k}|^2 \pm 2\sqrt{(\epsilon_{\mathbf{k}}^2 + m^2)\mu^{2}}}$, where $m$ is the AFM order parameter as defined in Eq. \ref{eq:coupling1}.
The coupled gap equations are:
\begin{subequations}
\begin{align}
\frac{6}{J} &=\sum_{\substack{l=\pm1\\\mathbf{k}}} \frac{1}{N_{\mathbf{k}}} \frac{1+l\frac{\mu^2}{\sqrt{(\epsilon_{\mathbf{k}}^2 +
m^2)\mu^{2}}}}{\sqrt{(\epsilon_{\mathbf{k}}^2 + m^2 + \mu^{2} +
|D_\mathbf{k}|^2 + 2 l \sqrt{(\epsilon_{\mathbf{k}}^2 + m^2)\mu^{2}}}} \\
\frac{12}{D} &= \sum_{\substack{l=\pm1\\\mathbf{k}}} \frac{1}{N_{\mathbf{k}}}
\frac{|\sum_{\mathbf{a}}f_{\mathbf{a}}|^2}{\sqrt{(\epsilon_{\mathbf{k}}^2
+ m^2 + \mu^{2} + |D_\mathbf{k}|^2 + 2 l \sqrt{(\epsilon_{\mathbf{k}}^2
+ m^2)\mu^{2}}}} \label{eq:GE-d}
\end{align}
\label{eq:GE}
\end{subequations}
where $N_{\mathbf{k}}$ is the number of $\mathbf{k}$ points in the summation.
Here, we use $l=+/-1$ to distinguish the valence/conduction bands in the original Dirac dispersion.
In the $m=0$ limit, if we only consider the $l=-1$ part of
Eq. \ref{eq:GE-d}, it reduces to Eq. \ref{eq:GapEq_sc}.
The appearance of the $l=1$ is because here we consider
intra-valence band pairing. At finite chemical potential
(which is the regime where SC emerges), the $l=1$ term always has a larger denominator and is therefore less important.
\begin{figure}
\begin{centering}
\includegraphics[width=0.45\textwidth]{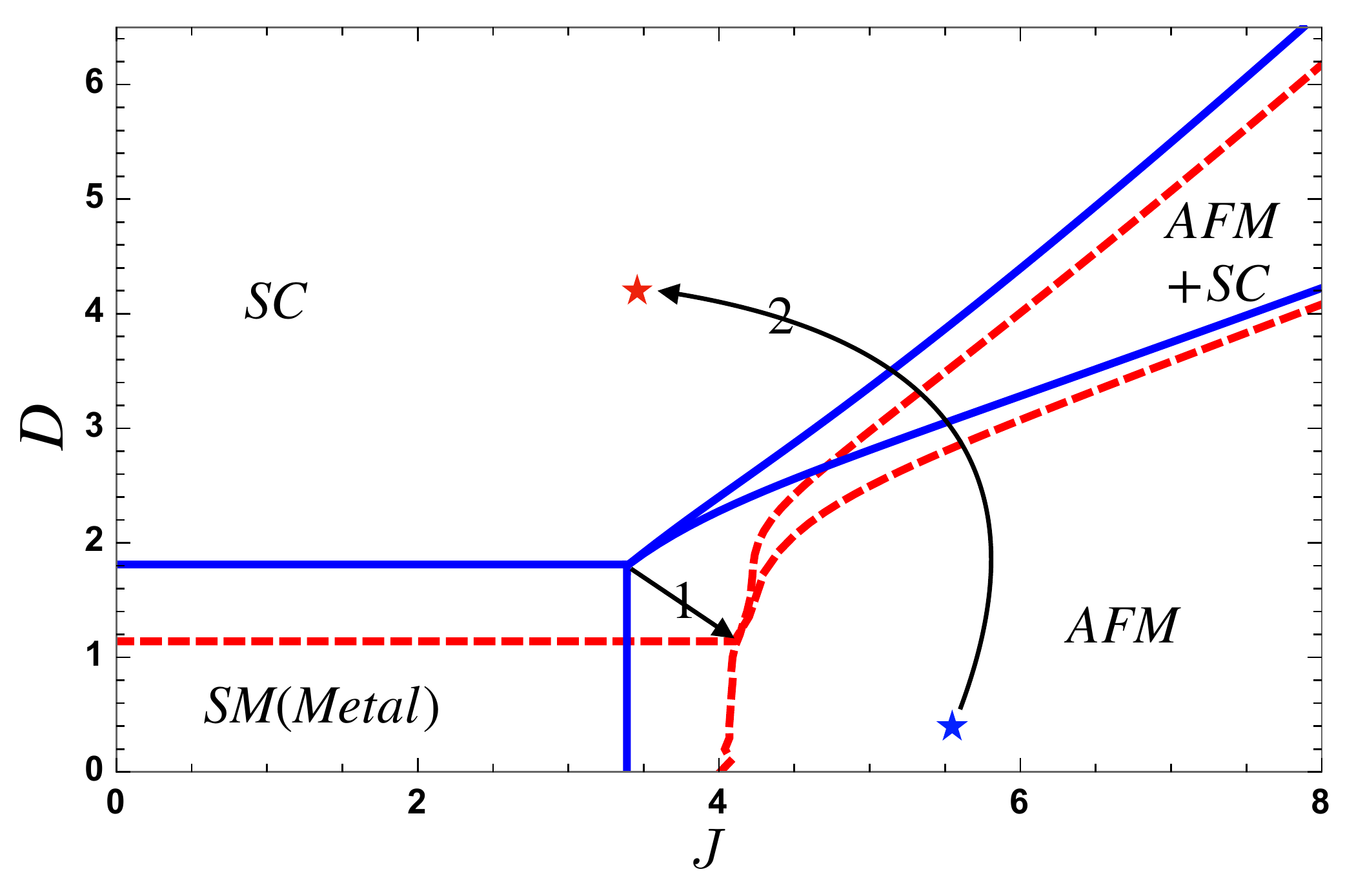}
\end{centering}
\caption{Mean field phase diagram for $\mu=0$ (blue solid line) and $\mu=0.3$ (red dashed line). The system is (semi-)metallic in the
weak coupling region. In the strong $J$ limit, it becomes AFM ordered
and in the strong $D$ limit, it is superconducting.
Arrow 1 shows qualitatively the increase in $J_c$ and decrease in $D_c$ upon doping, while arrow 2 shows the transition from AFM state to SC observed experimentally when tuning chemical potential.}
\label{fig:fig3}
\end{figure}

Figure \ref{fig:fig3} shows the mean field phase diagram.
The blue solid line is for $\mu=0$ while the red dashed line is for $\mu=0.3$. In both cases, the system is (semi-)metallic in the weak-coupling region. In the strong $J$
limit, it is AFM ordered while in the strong $D$ limit, it is
superconducting, as expected.
The opposite trend (indicated by arrow 1 in Fig. 3) of $D_c$ (which decreases) and $J_c$ (which increases) with doping implies that the competition between the two orders favours SC at higher carrier densities, and can be understood as follows: AFM order opens a gap at
the Dirac point.
Upon doping, the energy gain of AFM state becomes smaller,
which means a larger critical interaction is required.
On the other hand, SC opens a gap at Fermi surface, a larger Fermi surface at finite doping
indicates a smaller $D_c$. The gaps opening by AFM and SC order are shown in Fig. 1(c). This behavior is consistent with the effect of doping seen experimentally where doping drives tBG from an insulating to a SC state at low temperature: At half-filling, the tBG is insulating~\cite{cao2018correlated}, which our model indicates should correspond to an AFM insulating ground state. Starting from the AFM phase at $\mu=0$ in the diagram of Fig. 3, our results show that its stability is progressively reduced by adding more carriers, and ultimately replaced by a SC ground state beyond a critical doping. Note that the AFM-SC transition is not simply a consequence of the doping dependence of $J_c$ and $D_c$. In addition to that, $\frac{J}{t}$ is expected to decrease with doping~\cite{zhang1988renormalised} and enhance magnetic fluctuations. These, in turn, increase $V$ according to Eq. 4 further facilitating that phase transition. The transition from AFM insulating state to SC state is indicated by arrow 2 in Fig. 3.
%

%
\begin{figure}[htbp]
\begin{centering}
\includegraphics[width=0.48\textwidth]{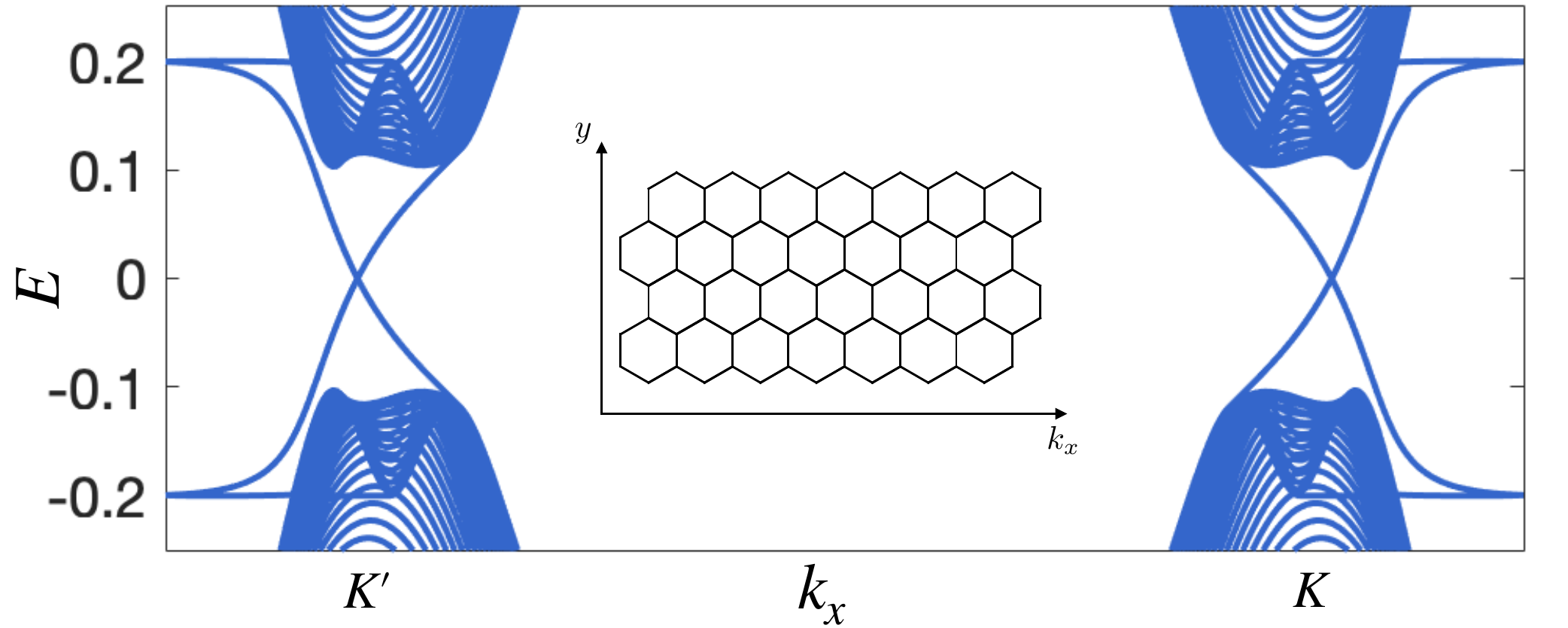}
\caption{tBG supports chiral Majorana modes in cylinder geometry. There is one chiral Majorana mode for each valley, which is due to the
$p+ip$ pairing around each Dirac point. The inset shows the geometry we
use. We take periodic boundary condition and open boundary condition in $y$ direction.}
\end{centering}
\label{fig:fig4}
\end{figure}
%

\emph{Chiral Majorana modes}.
It is known that the topology of the chiral $d+id$ pairing is non-trivial
because its Chern number is 2. As a result, there should be chiral
Majorana modes localized at the edge in a finite size system.
To show explicitly the existence of chiral Majorana modes, we studied a finite
(one direction) system with a cylinder geometry as can be seen in the inset of 
Fig. 4. We take a periodic boundary condition along the $x$ direction
and an open boundary condition along the $y$ direction. 
By diagonalizing the finite size Hamiltonian,
we obtain the spectrum in Fig. 4.
For each Dirac cone, there is one chiral edge mode at each edge.
This result is consistent with the observation that the pairing
has a $p+ip$ feature around each Dirac cone.
Near each edge, there are two chiral edge states in total,
reflecting the $C=2$ nature of the $d+id$ pairing.
The existence of these chiral Majorana modes can be detected in
transport experiment~\cite{he2017chiral}.


\emph{Acknowledgements}.
We thank Nimisha Raghuvanshi for helpful discussions. 
This work is supported by the Singapore Ministry of Education
MOE2017-T2-1-130 and MOE2017-T2-2-140.
\bibliographystyle{unsrt}
\bibliography{references}

\end{document}